\documentclass[aps,prl,preprint,superscriptaddress,showpacs]{revtex4-1}

\usepackage{longtable}
\usepackage{graphicx}
\usepackage{subfigure}
\usepackage{dcolumn}
\usepackage{bm}
\usepackage{amsmath}
\usepackage{amssymb}
\usepackage{color}

\begin{document}

\title{Kubo spins in nano-scale aluminum grains: A muon spin relaxation study.}

\author{Nimrod Bachar}
\email[]{nimrod.bachar@unige.ch}
\affiliation{Department of Quantum Matter Physics, University of Geneva, CH-1211 Geneva 4, Switzerland}

\author{Aviad Levy}
\affiliation{Raymond and Beverly Sackler School of Physics and Astronomy, Tel Aviv University, Tel Aviv, 69978, Israel}

\author{Thomas Prokscha}
\affiliation{Laboratory for Muon Spin Spectroscopy, Paul Scherrer Institute, 5232 Villigen PSI, Switzerland}

\author{Andreas Suter}
\affiliation{Laboratory for Muon Spin Spectroscopy, Paul Scherrer Institute, 5232 Villigen PSI, Switzerland}

\author{Elvezio Morenzoni}
\affiliation{Laboratory for Muon Spin Spectroscopy, Paul Scherrer Institute, 5232 Villigen PSI, Switzerland}

\author{Zaher Salman}
\affiliation{Laboratory for Muon Spin Spectroscopy, Paul Scherrer Institute, 5232 Villigen PSI, Switzerland}

\author{Guy Deutscher}
\affiliation{Raymond and Beverly Sackler School of Physics and Astronomy, Tel Aviv University, Tel Aviv, 69978, Israel}

\date{\today}

\begin{abstract}

We report muon spin relaxation rate measurements on films composed of aluminum grains having a size of a few nm, with a large energy level splitting of the order of 100~K. The films range from weakly metallic to insulating. In the insulating case the low temperature relaxation rate is consistent with the presence of single electron spins in grains having an odd number of electrons. The relaxation rate temperature dependence follows an activation law having an energy scale in agreement with the average level splitting. In weakly metallic films the relaxation rate is smaller and decreases faster with temperature. Overall our observations are in line with the presence of a Kubo spin in Al nano-size grains due to quantum size effects.    
       
\end{abstract}

\pacs{76.75.+i, 74.81.Bd}

\keywords{Muon Spin Relaxation; Kubo Spin; Granular Aluminum}

\maketitle

Several years ago Kubo~\cite{Kubo1962} predicted that small metallic particles having an odd number of electrons must bear an inherent magnetic moment because of the unpaired electron at the highest occupied energy level. The Kubo spin can be observed under two conditions. First of all, the thermal energy $k_B T$ must be small compared to the energy level splitting ${\delta}E$. For a particle radius of 5~nm this splitting is of the order of 1~K. Second, charge fluctuations in the particles must be restricted, as hopping in or out of an electron will erase the Kubo spin effect. In order to maintain charge localization in an ensemble of particles, the electrostatic charging energy must be much larger than the thermal energy. Kubo observed that for a particle of radius $R~=$~5~nm the charging energy $e^{2}/2R$ is much larger than the level splitting, and that this will still be the case if electrostatic interaction between neighboring particles is taken into account. Therefore, the energy level splitting sets the temperature scale where the Kubo spin can be observed. However, Kubo did not take into account the possibility of inter-grain electron tunneling, namely his theory strictly applies only to insulating samples. The effect of inter-grain tunneling on level broadening was studied by Kawabata~\cite{Kawabata1977}. In addition, as studied in detail in previous works~\cite{Florens2003}, a renormalization of the Coulomb blockade energy as a function of inter-grain conductance is expected. The effective Coulomb blockade energy decreases exponentially with inter-grain conductance and will eventually become smaller than the level splitting when inter-grain coupling increases.

\par 

A review of experimental attempts at observing the Kubo spin, in e.g. indium nano particles, has been presented by Perenbum \textit{et al.}~\cite{Perenboom1981,Perenboom1981b}. Despite the compelling prediction, Kubo spins were not so far confirmed experimentally. On the other hand, the presence of magnetic moments in weakly metallic granular Al thin films was inferred recently from magneto resistance measurements~\cite{Bachar2013} and confirmed by $\mu\text{SR}$ spectroscopy~\cite{Bachar2015}. We note that these granular Al films are superconducting at low temperatures, showing substantial inter-grain electron tunneling. The exact origin of these moments coexisting with enhanced superconductivity could not be completely clarified.     

\par

Here we present detailed and systematic $\mu\text{SR}$ measurements performed from room temperature down to 5~K on a series of granular Al films, ranging from insulating to weakly metallic. We show that the value and temperature dependence of the electronic relaxation rate in the insulating regime is fully consistent with the predictions of Kubo. The relaxation rate is smaller and vanishes more quickly with temperature in weakly metallic films. This behavior can be explained by the strongly renormalized Coulomb blockade energy, predicted theoretically.

\par 

Granular Al films were prepared by evaporation of clean (99.999\%) Al pellets in a partial pressure of oxygen. Most of the films were evaporated on 20~mm by 10~mm glass substrates that were held at liquid nitrogen temperature during the evaporation. This growth method has been proved to obtain films having a narrow grain size distribution centered around 2~nm diameter, with non-significant resistivity dependence of the distribution above 100~$\mu\Omega~cm$~\cite{Deutscher1972,*Deutscher1973,Deutscher1973a}. Some films were grown on substrates held at room temperature, lending 3~nm grains. With the increase of the partial pressure of oxygen, the insulating coating layer of the grain is becoming thicker and the resistivity increases due to the lower coupling between neighboring grains~\cite{Shapira1982}. A list of the films measured in this work and their main properties is given in Table~\ref{tb:Tab1}.

\begin{center}
    \begin{table}
        {\small
        \hfill{}
        \begin{tabular}{|l|l|c|c|c|}
        \hline
        Sample 	&	$\rho_{RT}$ 		& $d$ 	& $T_{c}$ 	& $E_{a}$ 	\\
        		&	$(\mu \Omega~cm)$ 	& $nm$ 	& $K$ 		& $K$ 		\\
        \hline
        \hline
        S140 	& 140 		& 2 	& 2.8 	& 55$\pm$4 \\ 
        S380 	& 380 		& 2 	& 3.1 	& 38$\pm$10 \\ 
        S1200 	& 1 222 	& 2 	& 3 	& 108$\pm$15 \\ 
        S1800 	& 1 816 	& 2 	& 2.9 	& 88$\pm$16 \\ 
        S10K 	& 9 440 	& 2 	& 1.8 	& 102$\pm$8 \\ 
        S100K 	& 100 000 	& 3 	& -- 	& 125$\pm$11 \\ 
        \hline
        \end{tabular}}
        \hfill{}
        \caption{Main characteristics of samples measured in this work. The room temperature resistivity $\rho_{RT}$ was measured for each sample. $T_c$ is estimated from the known SC phase diagram of granular Al thins films~\cite{Bachar2013}. The activation temperature $E_{a}$ was obtained from a fit of the muon spin relaxation rate, $\lambda$, experimental data to Eq.~\ref{eqn:ActLaw} using all temperature points.}
        \label{tb:Tab1}
    \end{table}
\end{center}

Zero field muon spin relaxation measurements in granular Al films having resistivities from 140~$\mu\Omega~cm$ to 100~000~$\mu\Omega~cm$ were conducted using low energy muon (LEM) spectroscopy~\cite{Morenzoni2000} at the Swiss Muon Source on the $\mu$E4 beamline~\cite{Prokscha2008} at the Paul Scherrer Institute in Switzerland. The implantation energy used in these measurements, typically 8 keV, was tuned using Trim.SP Monte Carlo simulations~\cite{Morenzoni2002} and chosen such that the muons are stopped in the middle of the 100~nm thick films. 

\par 

The polarization of the muon ensemble implanted in the samples, $P(t)$, is measured via detection of the emitted decay positron intensity as a function of time after thermalization. The time evolution of $P(t)$ is presented in the form of the asymmetry $AP(t)$ between two positron detectors, e.g. left and right orientations, where A is a constant determined by the muon decay properties and the geometry of the $\mu\text{SR}$ spectrometer.

\begin{figure}
    \begin{center}
        \includegraphics[width=0.8\columnwidth]{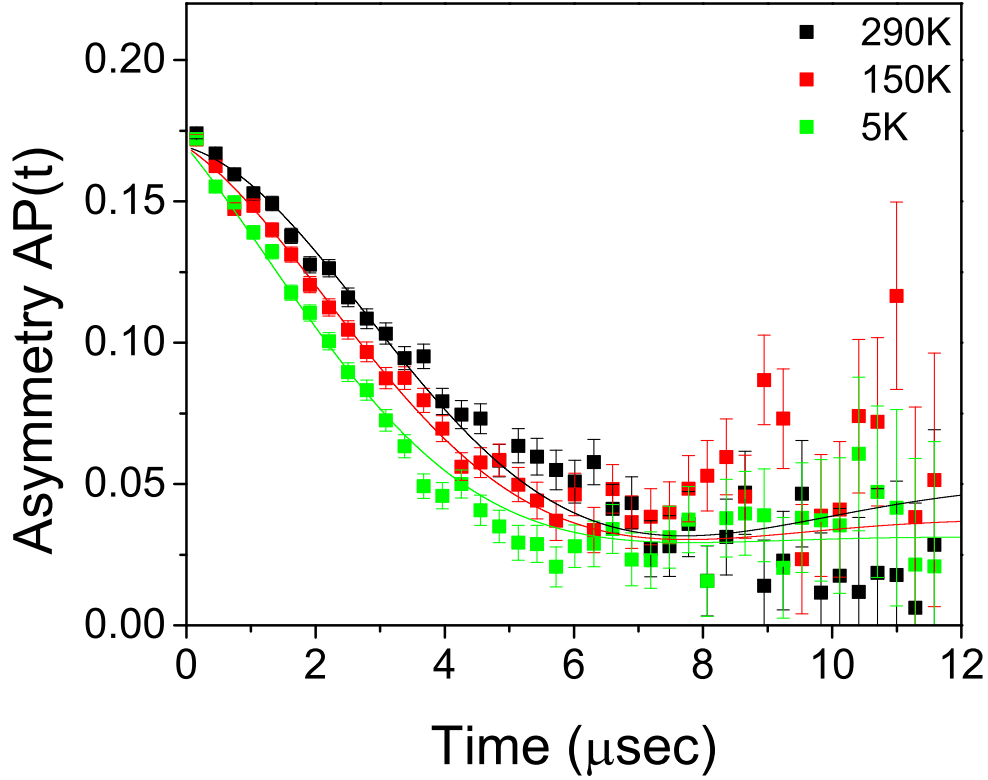}
        \caption{Time evolution of the asymmetry polarization signal (symbols) at selected temperatures for insulating granular Al sample S100K with room temperature resistivity of $100 000~~mu\Omega~cm$ with the corresponding fit (line) according to Eq.~\ref{eqn:KT}.}
        \label{fig:fig1}
    \end{center}
\end{figure}

Figure~\ref{fig:fig1} shows the asymmetry signal, $AP(t)$ measured in an insulating sample having a room temperature resistivity of 100~000~$\mu\Omega~cm$ (S100K). The figure shows data taken at of 290~K, 150~K and 5~K under zero field.

\par 

The shape of the asymmetry curves is clearly temperature dependent. At room temperature the curve shows at short times a negative curvature, while at low temperatures it gradually becomes more exponential-like. This change in shape with temperature indicates the presence of two distinct relaxation mechanisms. The first is due to the magnetic field produced by the Al nuclear moments, which is temperature independent. Therefore the temperature dependence of $AP(t)$ is due to a second mechanism which we attribute to fluctuating electronic magnetic moments present in the system. It is weak at room temperature due to fast fluctuations, but becomes comparable to the nuclear rate at low temperatures as the fluctuations slow down gradually.

\par 

In principle, muon diffusion can also produce a temperature dependent relaxation rate. Such a behavior has been observed in very clean Al samples~\cite{Kehr1982}. In our films, the smallness of the grains and the oxide coating of the Al grain acting as a barrier prohibits the muon diffusion process. Such a suppression has been observed also in previous measurements of granular~\cite{NoteSupp1} and of sputtered Al films~\cite{Morenzoni2002}.

\par 

To account for the two relaxation processes, the data was fitted with the Kubo-Toyabe function multiplied with an exponential function::
\begin{equation}\label{eqn:KT}
    AP(t)=A_{1}[\frac{1}{3}+\frac{2}{3}(1-\sigma^{2}t^{2})e^{-\frac{1}{2}\sigma^{2}t^{2}}]e^{-{\lambda}t} + A_{0}
\end{equation}
where the temperature independent relaxation rate $\sigma$ is attributed to the Gaussian distribution of the local magnetic field due to the nuclear moments of Al, and the temperature dependent relaxation rate $\lambda$ is caused by fluctuating electronic magnetic moments. The additional $A_{0}$ term accounts for a temperature independent background contribution from the sample backing plate and cryostat. 

\par 

According to Eq.~\ref{eqn:KT} the shape of $AP(t)$ changes continuously with the ratio $\lambda/\sigma$~\cite{NoteSupp1}. For small values of this ratio it starts out with zero slope and a negative curvature, goes through a minimum at $\sigma t \approx 1.73$, and rises back up to a value of 1/3. For values of $\lambda/\sigma$ of the order of one it starts down linearly, goes through a shallow minimum and settles back up to a value smaller than 1/3. The transition between these two regimes is clearly seen in the data shown in Fig.~\ref{fig:fig1}. From this data it follows that at room temperature the muons spin relaxation/depolarization is dominated by Al nuclear moments and that at low temperatures both electronic and nuclear moments contribute almost equally to the muon spin relaxation.

\begin{figure}
    \begin{center}
        \includegraphics[width=0.8\columnwidth]{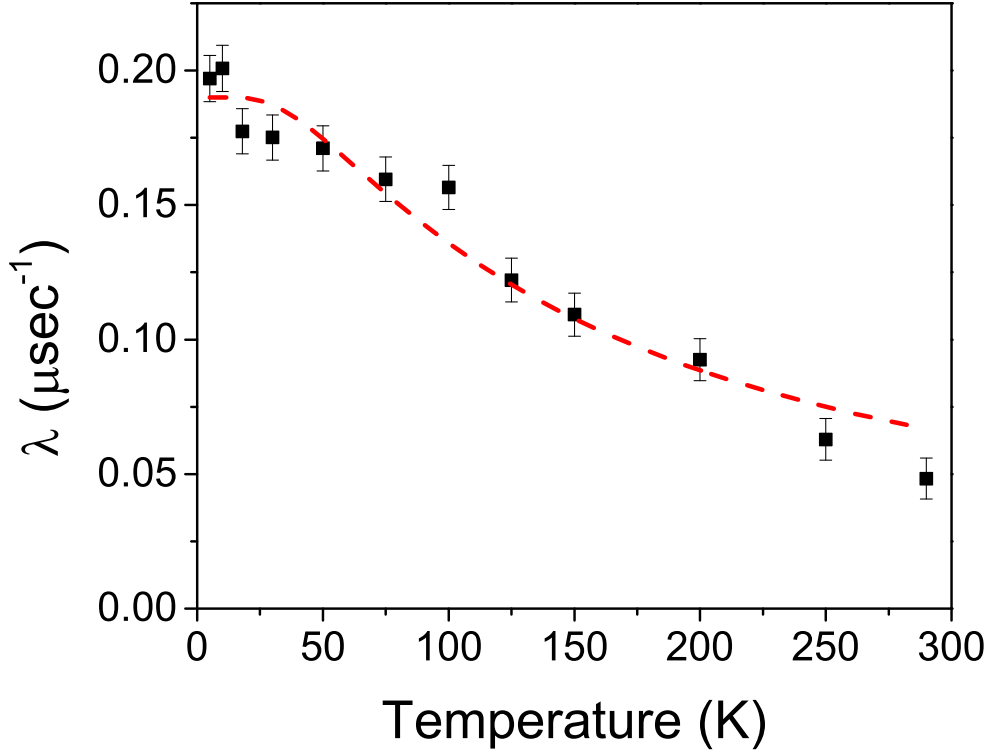}
        \caption{Temperature dependence of the muon spin relaxation rate of electronic origin $\lambda$ for sample S10K. The red dashed line follows Eq.~\ref{eqn:ActLaw} with $E_{a}\approx 125~K$ and $\lambda(0) \approx 0.19~\mu \textrm{sec}^{-1}$.}
        \label{fig:fig2}
    \end{center}
\end{figure}

The temperature dependence of $\lambda$ obtained from the fitted asymmetry curves for this insulating sample is shown in Fig.~\ref{fig:fig2}, together with a fit to the activation law:
\begin{equation}\label{eqn:ActLaw}
    \lambda(T)=\lambda(0)\left[ 1 - e^{-\frac{E_a}{k_B T}}\right] 
\end{equation}
This fit gives an activation energy $E_a$ of about 10~meV (See Table~\ref{tb:Tab1}). This value is much smaller than the electrostatic charging energy, which is 80~meV for isolated grains and estimated to be 30~meV when taking into account the electrostatic interaction between neighboring grains~\citep{Abeles1977}. Since in this insulating sample electrons are localized, there is no renormalization of the charging energy due to inter-grain electron tunneling. We therefore ascribe the measured activation energy to the average energy level splitting rather than to the charging energy. Taking into account that the Fermi energy of Al is 11.7~eV, we have thus about 1170 electronic levels per grain. Assuming the free electron model without the spin weight~\cite{Kubo1962} where $E_F/{\delta}E = 3ZN/4$ and that each Al atom contributes 3 electrons, i.e. $Z=3$, we have $N=520$ Al atoms per grain, which gives a spherical grain diameter of 2.5~nm. This is in excellent agreement with the average grain size of 3~nm measured in high resistivity Al granular films\cite{Deutscher1972,*Deutscher1973}.

\par 

At low temperatures $\lambda$ reaches a value of $0.197{\pm}0.0086~\mu \textrm{sec}^{-1}$. By comparing this value to that obtained for $\sigma=0.226{\pm}0.0046~\mu \textrm{sec}^{-1}$, one can calculate the fraction of Al atoms contributing to the electronic scattering~\cite{NoteSupp1}. It is of about one in 1000. We reach the conclusion that there is on the order of one Kubo spin for two grains, as expected~\cite{Kubo1962}.

\par 

\begin{figure}
    \begin{center}
        \includegraphics[width=0.8\columnwidth]{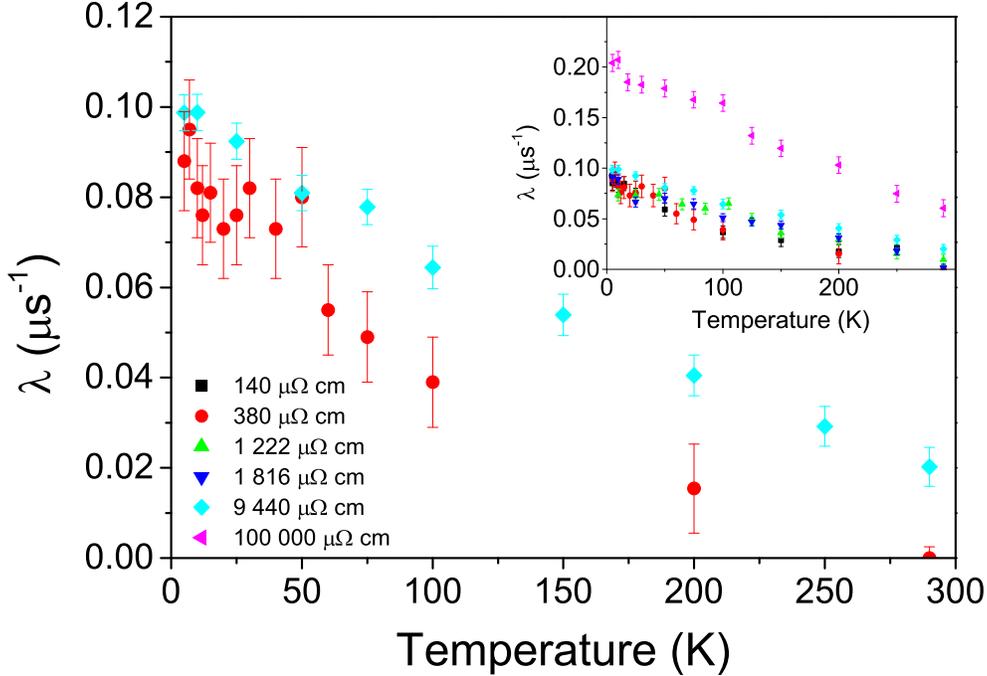}
        \caption{Temperature dependence of the muon spin relaxation rate of electronic origin $\lambda$ for samples S380 and S10K. Inset shows all samples measured in this work.}
        \label{fig:fig3}
    \end{center}
\end{figure}

We have also re-examined the electronic relaxation rate in weakly metallic samples having normal state resistivities in the range from 140~$\mu\Omega cm$ to about 10~000~$\mu\Omega cm$. At first look they seem to overlap but a closer examination reveals that there are differences amongst these metallic samples, particularly at high temperatures (Fig.~\ref{fig:fig3}). While for all samples the rate decreases as the temperature increases, in the highest resistivity one (S10K) it is still substantial at room temperature while for a lower resistivity one (S380) it is barely measurable. The value of the activation energy obtained from fits to Eq.~\ref{eqn:ActLaw} goes down progressively as the resistivity reduces, falling down to 30-50~K for the most metallic samples (see Table 1). For sample S10K it is still close to the value obtained for the insulating sample. But for the low resistivity sample the activation energy is definitely lower than the level splitting.

For all superconducting samples the rate saturates at low temperatures to a value of about $0.09~\mu \textrm{sec}^{-1}$, from which we calculated in a previous work a free spin concentration of about 1 spin per 4 to 5 grains~\cite{Bachar2015}. This concentration is definitely lower than that of the insulating sample.

\par 

The interpretation we propose is that the low activation energy in the metallic regime is the strongly renormalized electrostatic charging energy, rather than the energy level splitting. This re-normalization is known to decrease exponentially the charging energy of a quantum dot as a function of the conductance of the contacts that couple it to two bulk leads. For weak contacts:
\begin{equation}\label{eqn:EcRenorm}
    E_{C}^{\ast}=2\pi\alpha_{t}E_{C}e^{-\pi^{2}\alpha_{t}}
\end{equation}
Here $\alpha_{t}$ is the bare conductance of the contacts normalized to the universal conductance $G=e^{2}/h$. From the room temperature resistivity of the insulating sample and the grain size 3~nm the bare inter-grain conductance is of the order of $1{\times}10^{-6}~\Omega^{-1}$ and the renormalization effect is small. For the metallic samples the bare conductance of the inter-grain contact is of the same order as the universal conductance and the renormalization effect is large. At some point the effective charging energy becomes smaller than the level splitting and becomes the dominant energy scale for the observation of the Kubo spin. This apparently occurs when the normal state resistivity is about $10~000~\mu\Omega~cm$.

\par 

Another obvious difference with the behavior of the insulating sample is the smaller spin density at low temperatures. This is easily understood because Kubo spins are only present in grains where electrons are localized. The co-existence of localized and de-localized electrons in high resistivity metallic samples is known and leads to percolation effects in the superconducting state~\cite{Deutscher1980}.

\par 

Besides the Kubo spin, which is a volume effect, a surface effect may also be at the origin of magnetic moments~\cite{Bachar2015}. Since the oxide coating of the grain is not perfect, broken oxygen bonds or absorbed molecules on its surface~\cite{Lee2014,Wang2015} can result in a surface spin density. This may be one of the main causes for the phase decoherence effect in Quantum bit (Q-bit) devices fabricated using Superconducting Quantum Interference Device (SQUID) of either Al and Nb substance. Such spins were measured indirectly by the quantum flux noise $1/f$ in Q-bit devices~\cite{Sendelbach2008,Anton2013} and directly by Scanning SQUID magnetometry~\cite{Bluhm2009}. The experimental surface spin density was found to be about $4{\times}10^{17}~m^{-2}$. Recent theoretical works suggested that the origin of the $1/f$ flux noise is either due to strong coupling of free spins mediated by RKKY interaction from conduction electrons~\cite{Faoro2008} or due to weak anisotropic dipole-dipole coupling of spin clusters~\cite{Atalaya2014}. The latter was shown to explain the temperature dependence and geometry independence of the $1/f$ noise by taking into account the experimental surface spin density value of $4{\times}10^{17}~m^{-2}$.

\par 
 
However the temperature dependence of $\lambda$ that we observe cannot be easily explained by surface spins. Surface spins cannot hop between different grains thus do not have any reason to exhibit an activated temperature dependence, unlike Kubo spins which only appear at low temperatures when electrons become localized in each grain. We therefore propose that the volume Kubo spin rather than surface spins is at the origin of magnetic moments in granular Al in small metallic particles.

\par 

It is interesting to compare the activation energy scale obtained by $\mu\text{SR}$ in the metallic samples to that of the multi-level Kondo effect obtained from transport measurements. When the renormalized charging energy is smaller than the level splitting, the conductivity of a nano-dot is predicted to start out metallic-like at high temperatures, increasing as the temperature goes down, and to become insulating-like when the temperature reaches the renormalized charging energy~\cite{Florens2003}. This behavior was indeed observed in granular Al films having resistivites of 200 to 300~$\mu\Omega~cm$, with the downturn of the conductivity occurring at about 50~K~\cite{Moshe2017}. It is remarkable that this is indeed the value of the activation energy that we obtain from $\mu\text{SR}$ for similar resistivity values.

\par 

In conclusion, the $\mu\text{SR}$ results on granular Al films presented and analyzed here give for the first time a direct experimental evidence of the Kubo spin effect. As expected, this evidence is most compelling in insulating films. The activation energy obtained from the temperature dependence of the electronic relaxation rate, 125~K, is in excellent agreement with the average energy level splitting in an ensemble of metallic particles having the size seen by electron microscopy, and the number of spins per grain is of the order of one for a couple of grains. The value of the activation energy in metallic samples is reduced and is in agreement with the effective charging energy obtained from a multi-level Kondo analysis of the transport data for these films. 

\section{Supplementary material}

\begin{figure}
    \begin{center}
        \includegraphics[width=0.8\columnwidth]{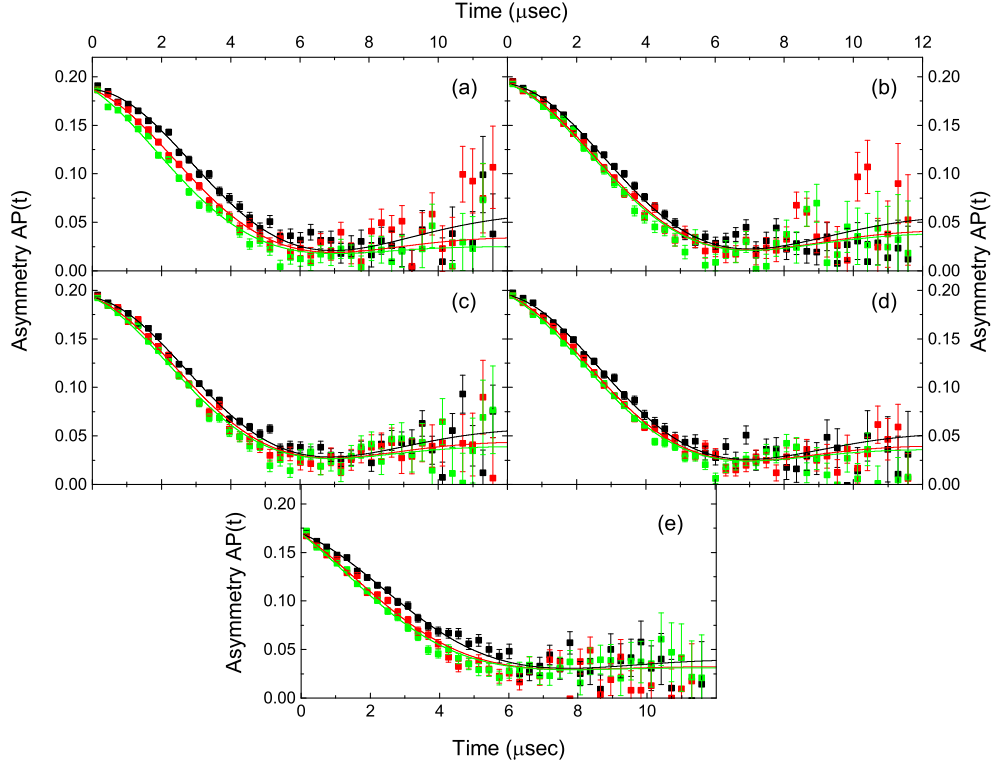}
        \caption{Summary of asymmetry curves at selected temperatures of 200~K, 50~K and 5~K for samples measured in this work and shown in Table~I of the main text in the following order: (a) $380~\mu\Omega~cm$ (b) $1~222~\mu\Omega~cm$ (c) $1~816~\mu\Omega~cm$ (d) $9~440~\mu\Omega~cm$ (e) $100~000~\mu\Omega~cm$.}
        \label{fig:app_fig1}
    \end{center}
\end{figure}

Figure~\ref{fig:app_fig1} shows a summary of selected asymmetry curves as were  measured in this work on samples of granular Al thin films (see Table~I of the main text). The data is shown for the same selected temperatures of 200~K, 50~K and 5~K. The muons were implanted in a typical energy of 8~keV and the data was collected while cooling down the sample and under zero magnetic field with a spin rotation angle of $-10^{\circ}$ with respect to the direction to the left detector. The data in Fig.~\ref{fig:app_fig1} obtained for the left and right set of detectors and is shown using a view packing of 1500 bins.

\begin{figure}
    \begin{center}
        \includegraphics[width=0.8\columnwidth]{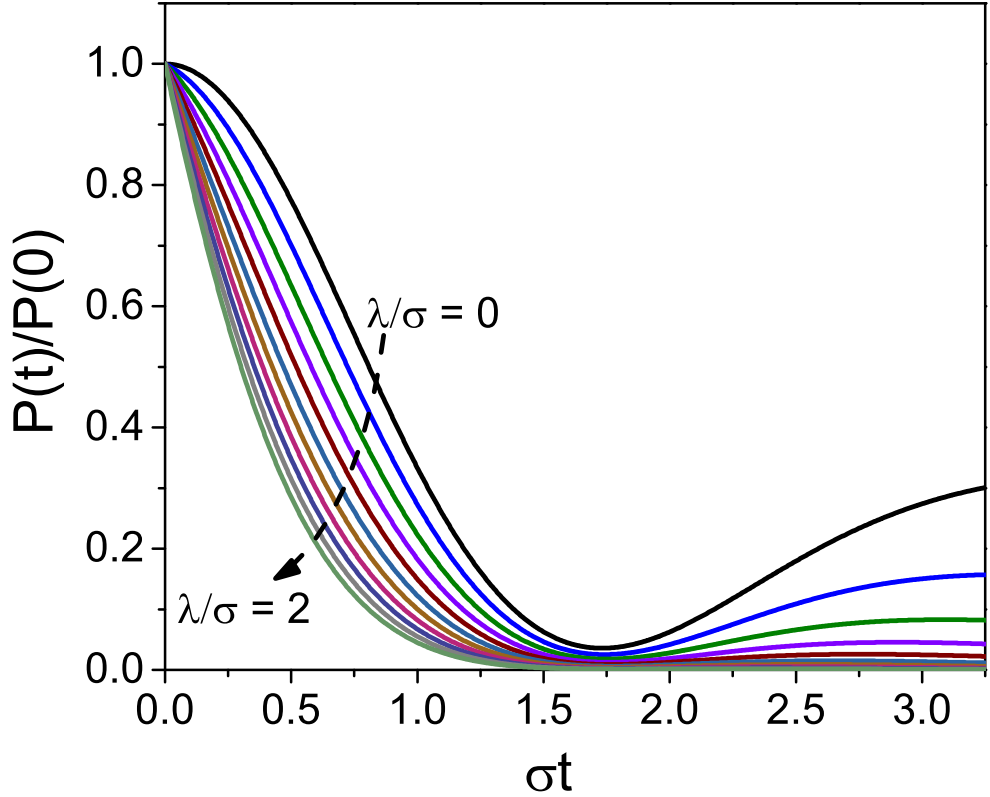}
        \caption{Simulation of the normalized asymmetry plot using the function given in Eq.~1 of the main text. The curves were generated for ratio of the electronic relaxation rate $\lambda$ and the nuclear relaxation rate $\sigma$ from 0 to 2 in steps of 0.2.}
        \label{fig:app_fig2}
    \end{center}
\end{figure}

The time evolution of the asymmetry curve can be described by the formula as given in Eq.~1 of the main text. As noted before it includes two terms which are attributed to the muon's relaxation rate: The first is the nuclear relaxation rate $\sigma$ which is attributed to the Al nuclear moment and is assumed to be temperature independent. The second is the electronic relaxation rate $\lambda$ which is attributed to electronic free spins in the sample and can show temperature dependency. Given the exponential contribution of each of these terms, we show in Fig.~\ref{fig:app_fig2} a simulation of the expected asymmetry time dependent signal given a zero time value of $AP(0)=1$. In a Kubo-Toyabe formula which includes only one relaxation rate term, i.e. $\lambda/\sigma=0$, the asymmetry signal starts from a its maximum value at $t=0$ with a zero slope and a negative curvature, reduces to a minimum at about $\sigma t = 1.73$ and asymptotically approaches 1/3 of its original value at longer times. However when $\lambda$ is included the time dependency of the asymmetry signal is altered from the typical single relaxation rate form. Although the minimum in $\sigma t$ is maintained, one can see that the signal at $t=0$ has already a negative nearly linear drop while the long time value is reduced dramatically. Such signatures in the asymmetry curves suggest that the contribution of the electronic relaxation rate is increased either by temperature or resistivity as can be clearly observed in our raw data of Fig.~\ref{fig:app_fig1}.       
    
Figure~\ref{fig:app_fig1} shows also the equivalent fit of the data to the Kubo-Toyabe equation with additional exponential decay (Eq.~1 of the main text) including both relaxation rate terms, $\sigma$ and $\lambda$. The asymmetry data was fitted in the range of $0.07~\mu \textrm{sec}$ to $10~\mu \textrm{sec}$ in order to exclude the short time decay rate coming from the Ni substrate and the lower signal-to-noise at long time. We used a packing of 50 bins for the above fit. We applied the fit on all available curves for one sample during one cool down. The parameters used for the fit are the nuclear relaxation rate, $\sigma$, the asymmetry zero time value for the sample, $AP_{s}$, and the background, $AP_{b}$, the sample geometrical factor $\alpha$ and the electronic relaxation rate $\lambda$. Before each temperature run and for each sample we performed a field dependent measurement at $B=50G$ and at room temperature in order to obtain the geometrical factor $\alpha$ to be used for fitting the zero field measurements. Aforementioned parameters except $\lambda$ were shared in our fit procedure resulting in common values for the entire temperature run of each sample. As a result a typical value of $\sigma \approx 0.25\pm0.01~{\mu}\textrm{sec}^{-1}$ was shown to be resistivity independent while only at the insulating sample this value changed to $\sigma \approx 0.22~{\mu}\textrm{sec}^{-1}$. As noted before~\citep{Bachar2015} such values are in between the expected values for pure Al and Al oxide. From the above fit procedure, we finally deduced the temperature dependent value of $\lambda(T)$ as shown in Fig.~3 of the main text.       

\par 

From the values of the electronic relxation rate at low temperatures we extracted the estimated concentration of free spins~\cite{Bachar2015}. For the metallic and superconducting samples ranging from $140~\mu \Omega cm$ to about $10~000~\mu \Omega cm$ we obtained a relaxation rate value of $\lambda\approx 0.09 \mu \textrm{sec}^{-1}$ which results in a free spins concentration of about 400~ppm. Considering the estimation that each grain holds about 520 Al atoms, we obtain a spin concentration of about 1 spin per 4 to 5 grains. The value of $\lambda\approx 0.19 \mu \textrm{sec}^{-1}$ for the highest resistivity sample of $\rho \approx 100~000~\mu \Omega cm$ is about twice the ones of the lower resistivity samples. We obtain a concentration of about 800~ppm or 1 spin per two grains in the insulating side, as expected in the case of a Kubo spin in isolated finite size metallic grains. 

\par 

Muons diffusion can in principle induce a temperature dependence of the depolarization rate due to the nuclear moments. This diffusion was studied down to the lowest temperatures (30~mK) and found to take place in very pure Al with less than 1 ppm overall impurities and RRR$>$10000. Upon adding impurity atoms in the few to $>$1000~ppm level (see e.g. Ref.~\cite{Kehr1982}), due to trapping-detrapping processes a non-monotonic relaxation rate vs. temperature behavior was found with maxima of the relaxation rate around 10-50~K. The maxima correspond to the muon being static. The trapping rate was found by Kehr~\textit{et al.}~\cite{Kehr1982} to be proportional to the impurity concentration.

\par 

By contrast our samples consist of very small Al grains (about 2~nm) with a solid oxide barrier between them, which will impede the muons diffusion. For a rough estimate of the equivalent impurity level in nanograin Al, we assumed an oxide layer on a 2~nm Al sphere. This gives about 50 promil impurity concentration. The effective number of trapping center can be even higher than that because an impurity can produce several (up to 100) trapping centers. Taking the trapping rates from Kehr~\textit{et al.}, we get a trapping time of 20~ns, which is even an upper limit so that we can safely assume that muon diffusion does not influence our results.

\par 

Therefore, muon diffusion can be neglected and the temperature independent Kubo-Toyabe rate $\sigma{\approx}0.25~{\mu}\textrm{sec}^{-1}$ that we observe corresponds to a static muon depolarization due to the nuclear moments in Al and O, and the monotonous exponential rate is of electronic origin.

\bibliographystyle{apsrev4-1}
\bibliography{Bachar_muSR_2018}

\end{document}